\documentclass[aps,prl,reprint,superscriptaddress]{revtex4-1}

\usepackage{graphicx,float,gensymb}

\pdfoutput=1

\begin{document}

\title{Ultrafast spin density wave transition in Chromium governed by thermalized electron gas}

\author{C. W. Nicholson}
\affiliation{Department of Physical Chemistry, Fritz-Haber-Institut of the Max Planck Society, Faradayweg 4-6, Berlin 14915, Germany}

\author{C. Monney}
\affiliation{Department of Physics, University of Zurich, Winterthurerstrasse 190, 8057 Zurich, Switzerland}

\author{R. Carley}
\altaffiliation{Present address: European XFEL GmbH, Notkestrasse 85, 22607 Hamburg, Germany}
\affiliation{Max-Born-Institut, Max-Born-Strasse 2A, 12489 Berlin, Germany}

\author{B. Frietsch}
\affiliation{Max-Born-Institut, Max-Born-Strasse 2A, 12489 Berlin, Germany}
\affiliation{Fachbereich Physik, Freie Universitaet Berlin, Arnimallee 14, 14195, Berlin, Germany}

\author{J. Bowlan}
\altaffiliation{Present address: CINT, Los Alamos National Laboratory, Albuquerque, New Mexico, USA}
\affiliation{Max-Born-Institut, Max-Born-Strasse 2A, 12489 Berlin, Germany}

\author{M. Weinelt}

\affiliation{Fachbereich Physik, Freie Universitaet Berlin, Arnimallee 14, 14195, Berlin, Germany}

\author{M. Wolf}
\affiliation{Department of Physical Chemistry, Fritz-Haber-Institut of the Max Planck Society, Faradayweg 4-6, Berlin 14915, Germany}

\date{\today}

\begin{abstract}
The energy and momentum selectivity of time- and angle-resolved photoemission spectroscopy is exploited to address the ultrafast dynamics of the antiferromagnetic spin density wave (SDW) transition photoexcited in epitaxial thin films of chromium. We are able to quantitatively extract the evolution of the SDW order parameter $\Delta$ through the ultrafast phase transition. $\Delta(T_{e})$ is defined by the transient temperature of the thermalized electron gas, $T_{e}$. The complete destruction of SDW order on a sub-100~fs time scale is observed, much faster than for conventional charge density wave materials. Our results reveal that equilibrium concepts for phase transitions such as the order parameter may be utilized even in the strongly non-adiabatic regime of ultrafast photo-excitation.
\end{abstract}

\maketitle

In ultrafast pump-probe measurements, the transfer of energy from the laser pulse can lead to the population of excited states \cite{Gierz2013, GrubisicCabo2015a, Sobota2012}, changes in magnetic ordering \cite{Radu2011, Frietsch2015} and even electronic or structural phase transitions \cite{Schmitt2008, Rohwer2011, Beaud2014, Waldecker2015}. In many cases a transient increase of the electronic temperature occurs which may be tracked, for example, by angle-resolved photoemission spectroscopy (ARPES) \cite{Gierz2013, Wang2012,Crepaldi2012a}.
An open question is still to what extent the electronic temperature alone can be said to govern ultrafast changes, particularly for phase transitions, due to the strongly non-adiabatic nature of pump-probe experiments and the possibility of exciting non-thermal electron distributions on short time scales. Such a description is further complicated in many correlated materials such as high-Tc superconductors, charge density waves (CDWs) and ferromagnets, in which lattice degrees of freedom play an important role. In the case of CDWs, the periodic motion of the atomic cores (phonons) can lead to a periodic opening and closing of the spectral gap \cite{Schmitt2008, Rettig2016, Yang2014, Papalazarou2012} independent of the temperature of the electronic system. In ultrafast demagnetisation of ferromagnets, a bottle neck for the transition is the transfer of angular momentum, which proceeds through the lattice \cite{Stohr2006} meaning that a hot electron system may be necessary, but is not sufficient to drive the system from one magnetic phase to another.
In contrast, the ordering in spin density waves stems directly from electronic correlations \cite{Overhauser1962} and thus offers an opportunity to study the dynamics of a phase transition in which the role of the lattice is minimised. This allows a more stringent test of the role played by the electronic temperature in driving materials from one phase to another under non-equilibrium conditions.

Cr famously undergoes a transition to an antiferromagnetic-SDW phase \cite{Fawcett1988}.
Although the SDW in Cr has been widely studied \cite{Fawcett1988,Schafer1999,Fawcett2000,Rotenberg2005,Rotenberg2008}, very few studies of the time-domain dynamics exist \cite{Hirori2003,Singer2015,Singer2015a}, none of which directly address the electronic structure. 

\begin{figure}
		\includegraphics[scale=0.3]{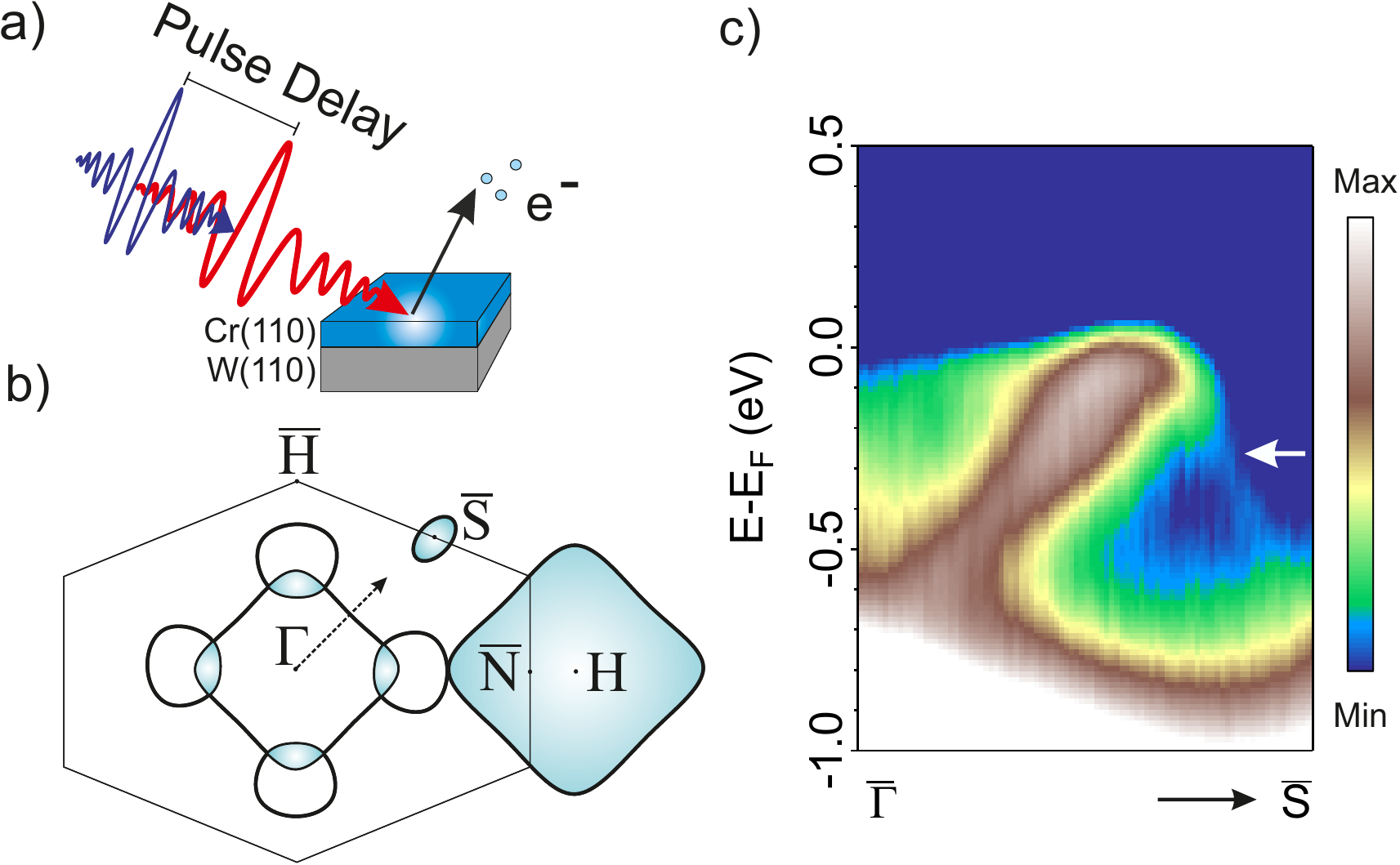}
		
	\caption{\small{(Color online) (a) Schematic pump-probe experiment of the Cr(110) thin film. (b) Schematic Fermi surface and surface Brillouin zone of Cr. Electron pockets are contained within solid black lines while hole pockets are colored cyan. The hexagon marks the (110) surface projected Brillouin zone. (c) Electronic structure of Cr(110) thin film measured along the dotted arrow in Fig~1b. The SDW band is marked by a white arrow.}} 
	\label{fig:Figure 1}

\end{figure}

In this Letter we report the first time-resolved ARPES investigation of the SDW transition in Cr. We directly observe the ultrafast disappearance and recovery of the SDW-derived electronic structure following pulsed infra-red excitation. The ultrafast dynamics of the electronic structure are simulated with a mean field model in order to disentangle intrinsic SDW dynamics from other non-equilibrium effects. We find that the order parameter of the SDW is governed by the transient electronic temperature, implying an intimate link between electronic temperature and spin ordering as the driving mechanism of the SDW in Cr. Our results demonstrate that equilibrium concepts can still survive on ultra-short time scales. 

\begin{figure}
		\includegraphics[scale=0.35]{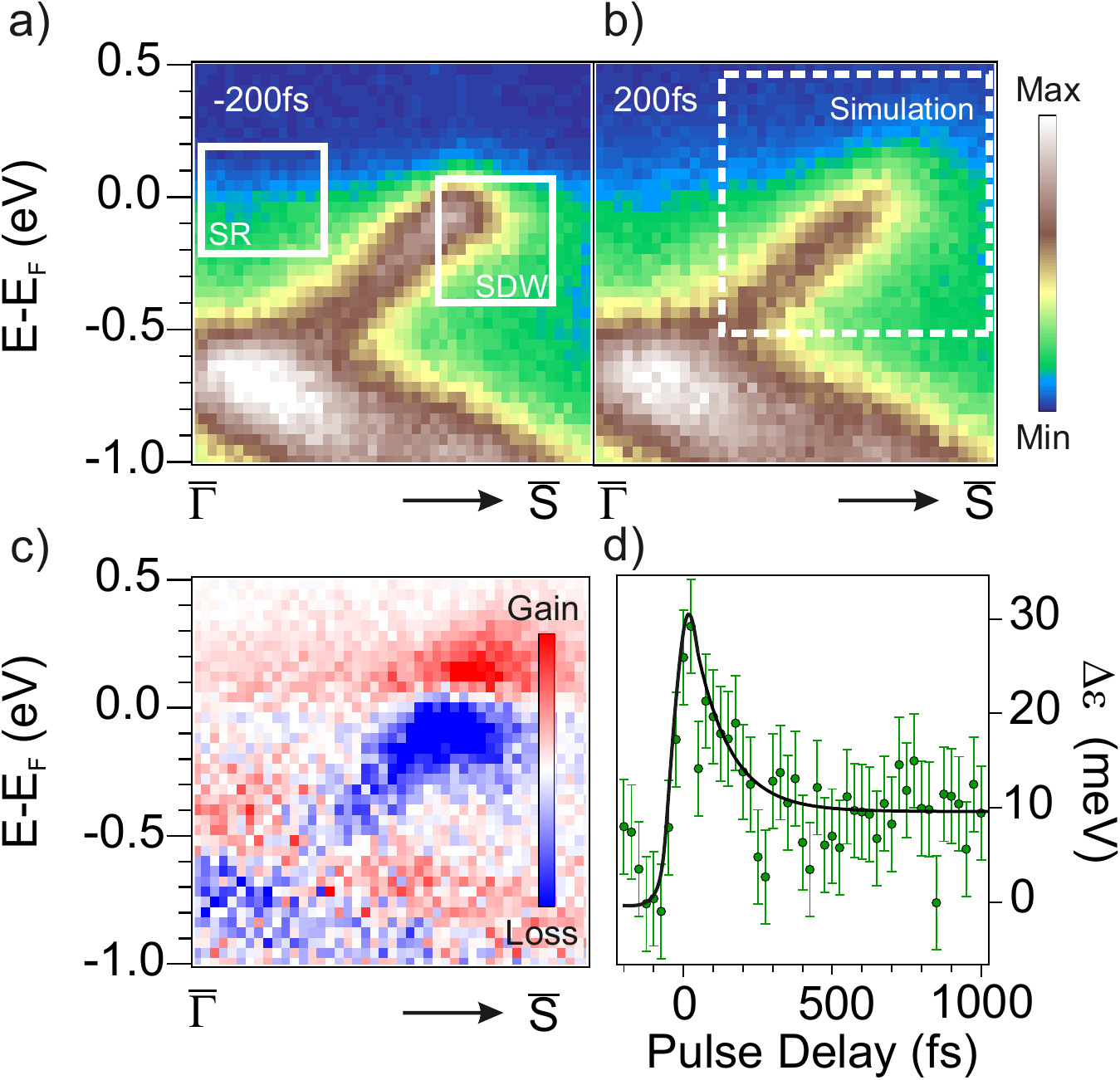}
		
	\caption{\small{(Color online) ARPES images obtained with 40~eV probe energy (a) in thermal equilibrium at a delay of -200~fs (XUV probe pulse before pump pulse) and (b) following optical excitation at 200~fs delay. The dotted box highlights the area considered in the simulation as presented in Fig.~4a. (c) Difference image between the images shown in (a) and (b). In the false color plot red and blue mark intensity increase and decrease of intensity respectively. (d) Transient rigid shift of the EDCs. The solid black line is a guide to the eye.}} 
	\label{fig:Figure 2}

\end{figure}

Cr(110) films were grown epitaxially on a clean W(110) crystal at room temperature at a pressure of 1x10$^{-10}$~mbar, and then annealed to 600\degree C. A film thickness of 7~nm was produced by reference to the LEED and ARPES phase diagram \cite{Rotenberg2005}. For the pump-probe measurements, shown schematically in Fig.~1a, a linear polarised 1.5~eV pump and 40~eV XUV (extreme ultraviolet) probe were utilised to collect snapshots of the electronic structure. A detailed description of our high-harmonic generation-based trARPES set-up is presented elsewhere \cite{Frietsch2013}. Due to the optical-absorption depth of 30~nm at the pump wavelength \cite{Rakic1998}, the entire thin film will be excited nearly homogeneously. The combined time resolution of the experiment was measured to be 130~fs. Unless otherwise stated, a fluence of 0.8~mJ~cm$^{-2}$ and an s-polarised pump were selected to maximise pump-probe signal while avoiding significant space charge. ARPES measurements were carried out with a hemispherical electron analyser (SPECS) at 100~K, well below the surface SDW transition temperature of 411~K \cite{Schafer1999}. All results presented here are from the band at the Fermi level $E_{\rm{F}}$ along the $\overline{\Gamma}$-$\overline{S}$ direction marked in Fig.~1b. 

Fig.~\ref{fig:Figure 1}c shows the electronic structure of Cr in the SDW phase measured with He I radiation. The data are plotted in a false colour log scale to highlight weak features. An electron band is seen to disperse towards the Fermi level ($E_{\rm{F}}$) from $\overline{\Gamma}$ towards the $\overline{S}$-point.
As the band approaches $E_{\rm{F}}$ it bends away and continues again to higher binding energies. This feature with weak spectral weight is a direct result of the SDW periodicity which results in a band renormalisation via back folding \cite{Rotenberg2005}.
The fact that we observe the renormalised dispersion in our data confirms the high quality of our Cr film preparation. A hole-like band with maxima at -0.7~eV binding energy is also derived from Cr states (not shown in Fig.~1c). 

The same spectrum obtained with pulsed XUV radiation is presented in Fig.~\ref{fig:Figure 2}a. As expected in the SDW phase the renormalised band is observed, though it is weaker than in the He-lamp measurements due to the lower energy resolution of the high harmonics source of 200~meV. Upon excitation with the pump laser pulse, the dispersion is observed to change. As shown for 200~fs after optical excitation, the dispersion linearly crosses the Fermi level, as in the paramagnetic phase \cite{Schafer1999, Rotenberg2005}, while spectral weight is removed from the renormalised band. These changes are highlighted by the difference image in Fig.~2c in which one clearly sees a strong reduction (blue) of spectral weight in the SDW band region, which is gained above the Fermi level (red). Additionally, between the SDW band and $\overline{\Gamma}$, intensity is gained above $E_{\rm{F}}$ due to the broadening of the Fermi-Dirac distribution of the metallic states from a surface resonance \cite{Rotenberg2008}.

Also visible in Fig.~2c at 200~fs is a rigid shift of the entire electronic structure towards higher kinetic energies. A shift of almost 30~meV is observed which is the same, within experimental errors, as the observed shift of the Fermi level. The dynamics of this shift over the time delay of the experiment are given in Fig.~2d, revealing a rapid peak followed by recovery to a new equilibrium value by around 500~fs. Such dynamics are not typical of probe-induced space charge effects. The origin of this shift may be due to out-of-equilibrium chemical potential shifts, or a change of the potential barrier at the sample surface induced by the transient electronic temperature. For simplicity we treat it purely phenomenologically in the following analysis. 

Energy Distribution Curves (EDCs) from the metallic surface resonance region (box ``SR'' in Fig.~2a) are presented in Fig.~3a at different pulse delays. By fitting a Fermi-Dirac distribution convolved with the instrumental resolution, the transient electronic temperature ($T_{e}$) may be extracted. We note that within our time resolution of 130~fs we always observe a thermal distribution of electrons. Fig.~3b displays the evolution of $T_{e}$ up to 1000~fs following excitation. After a rapid increase to 1200~K close to time zero, $T_{e}$ decreases to a new quasi-equilibrium value after 800~fs, where it is thermalized with the lattice. These results are corroborated by a simple two-temperature model fit to the data (see Supplementary Material for details \cite{SuppMat}).

\begin{figure}
		\includegraphics[scale=0.35]{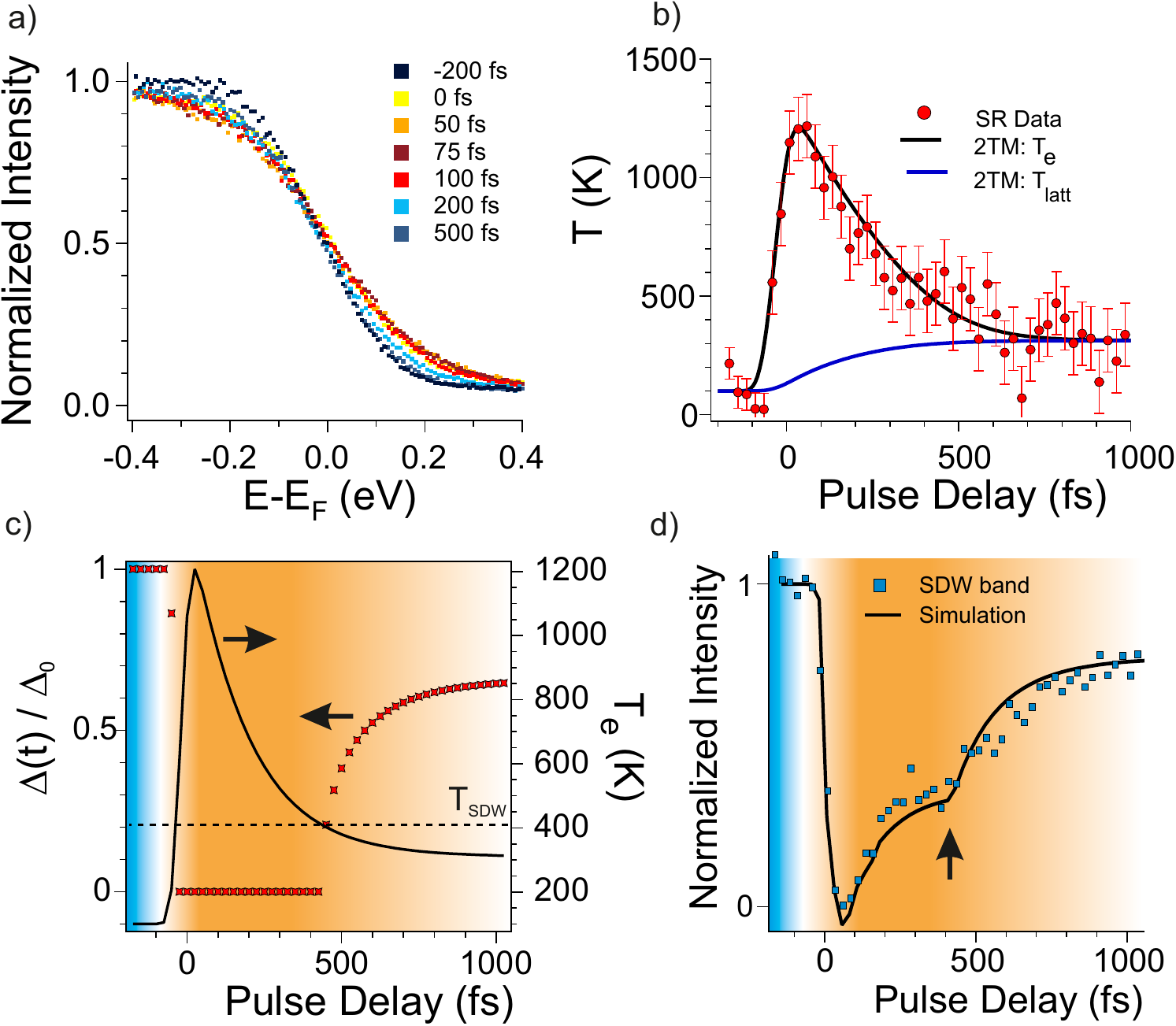}
		
	\caption{\small{(Color online) (a) Fermi distribution extracted from region SR indicated in Fig.~2a at various pulse delays. (b) Electronic temperature extracted from the Fermi distribution as a function of pulse delay. Solid black and blue lines are 2TM fits to the data for the electron and lattice temperature respectively (see main text). (c) Electronic temperature (solid black line) and associated order parameter (red markers) calculated as a function of pulse delay. The dotted line marks the SDW transition temperature. (d) Dynamics of the normalized SDW band intensity in the region marked SDW indicated in Fig.~2a for both the simulation (solid line) and experiment (markers). An arrow marks the point at which $\Delta$ transitions from zero to a finite value. The color background in (c) and (d) highlights the region where the SDW is in the ground (blue) and excited (orange) state.}} 
	\label{fig:Figure 3}

\end{figure}

In order to investigate the quantitative effect of $T_{e}$ on the SDW electronic structure, we utilize results from a mean field model, which may be applied to BCS superconductors and density waves \cite{Mahan2000,Gruner2000}, in order to describe the SDW as a renormalised dispersion with a gap. In such a model, the poles of the Green's function give the quasi-particle dispersion, which may be written as $E_{\pm}=\pm\sqrt{\epsilon(k)^{2}+\Delta^{2}}$, where $\epsilon$ is the bare band dispersion. This renormalised dispersion has two branches separated by a gap determined by $\Delta$. The temperature dependence of the SDW order parameter ($\Delta$) is given by

	\[\Delta(T_{e})=\Delta_{0}\sqrt{1-\left(\frac{T_{e}}{T_{C}}\right)^{2}}
\]

where $\Delta_{0}$ is half the gap measured by photoemission, and $T_{C}=411$~K is the surface phase transition temperature of the SDW \cite{Schafer1999}. The spectral weight in the renormalised bands is given by the coherence factors $u_{\pm}^{2}=(1+\epsilon/E_{\pm})/2$. Further details may be found in the Supplementary Material \cite{SuppMat}. In the experiment each time delay is associated to a particular $T_{e}$, hence we introduce time resolution into the simulation by producing snapshots of the electronic structure at the experimentally determined $T_{e}$. The rigid shift dynamics are added phenomenologically. The phonon (i.e. the lattice) temperature from a two-temperature model (2TM) calculation is also extracted \cite{SuppMat}. We note that the temperature of the phonons is always lower than the SDW transition temperature in our measurements (see Fig.~3b), and therefore does not play a dominant role in the phase transition dynamics.

The response of the order parameter to the transient $T_{e}$ is shown in Fig.~3c, and compared with $T_{e}$. In the region where $T_{e}>T_{C}$, $\Delta$ goes to zero, reflecting the closing of the SDW gap. Once $T_{e}$ drops below $T_{C}$, $\Delta$ becomes finite and tends towards a stable out-of-equilibrium value. The resulting simulated spectral response of the renormalised SDW band is shown in Fig.~3d as a solid black line. The region considered in the simulation is marked in Fig.~4a by a white box, which is equivalent to that used to extract the response in the data, marked in Fig.~2a. Following excitation two distinct regions of recovery are evident. The first rapid increase in intensity is due to the rigid band shift presented in Fig.~2d, which shifts the electronic structure relative to the fixed region of interest (ROI). In principle, such an effect may be removed by considering a ROI large enough to encompass both the bands plus the shift. However doing so means the ROI includes the region in which the Fermi distribution is transiently broadened and thus a transient change of intensity is also observed here due to changing thermal distribution. In our simulation it is found that the rigid shift dynamics and the dynamics of the Fermi distribution (modelled by the SDW density of states multiplied by the Fermi function) give the same qualitative response. The second recovery region is the result of the order parameter dynamics, as shown in Fig.~3c, which becomes finite at the point marked by the arrow in Fig.~3d, resulting in a shoulder in the recovery dynamics.

The simulated response is compared to the experimental data from the region of the SDW band (box ``SDW'' in Fig.~2a) and shows good agreement. 
The image from our simulation in Fig.~4a summarizes the effects taking place in the electronic structure. For a given pulse delay, the value of the order parameter is determined by $T_{e}$, which therefore defines both the size of the gap and the spectral weight of the renormalised band. In addition, $T_{e}$ drives the thermal broadening which results in an additional reduction of intensity for a fixed region of integrated electronic structure. As discussed, the rigid band shift can have the same effect.

In order to further test the idea that $\Delta$ is governed by $T_{e}$, we have carried out fluence dependent measurements of the SDW phase transition. If the time at which the SDW gap re-opens does indeed purely depend on $T_{e}$, it should vary with the laser fluence, as this changes the transient $T_{e}$ dynamics. Data sets at different fluences show similar shoulders to those observed in Fig.~3d, corresponding to the SDW gap re-opening \cite{SuppMat}. The time at which the gap re-opens is summarised in Fig.~4b as a function of absorbed fluence. Data sets for which $T_{e}>T_{SDW}$ is fulfilled for some delays show a gap re-opening; conversely when $T_{e}<T_{SDW}$ for all delays we observe only a single recovery time scale, resulting from the change of spectral weight due to the rigid band shift or the thermal distribution.
 
Below around 0.2~mJ~cm$^{-2}$ we find that the SDW is no longer driven into the paramagnetic phase, and that only a single relaxation time scale is observed. Above this fluence, the time taken to relax below $T_{C}$ increases with fluence. This corroborates the idea that upon ultrafast optical excitation the melting and recovery of the SDW is driven by a purely electronic mechanism and that we can indeed quantitatively track $\Delta$ through the ultrafast phase transition.

\begin{figure}
		\includegraphics[scale=0.4]{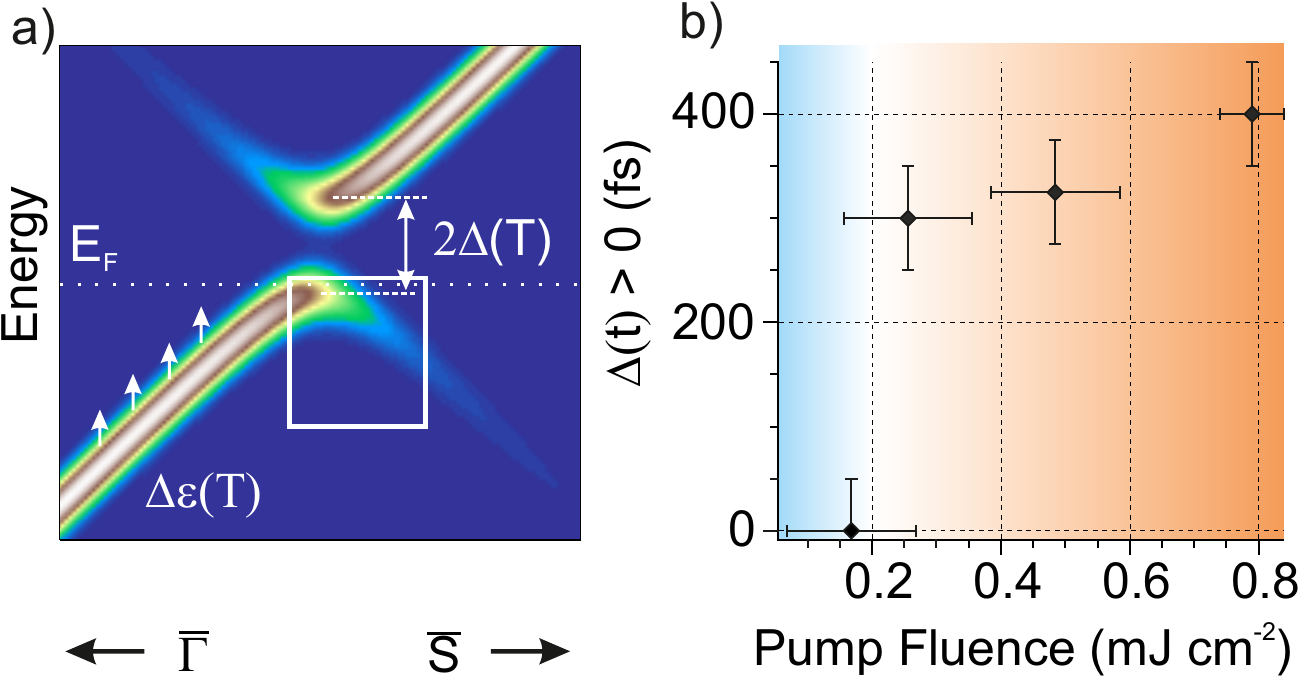}
		
	\caption{\small{(Color online) (a) Simulated electronic structure in the region highlighted in Fig.~2b. Following pump excitation both the order parameter and the rigid shift change transiently. No Fermi distribution has been applied in the image to highlight the two branches of the renormalised dispersion. (b) Absorbed fluence dependence of the time at which the SDW gap re-opens. The red region highlights the fluence range for which the SDW order parameter is zero for some delays, while in the blue region it remains finite for all delays.}} 
	\label{fig:Figure 4}

\end{figure} 

The disappearance of the SDW signature - the renormalised band - implies the electron gas no longer experiences the SDW spin-ordering potential, which results in the disappearance of the long-range spin ordering. Since AFM ordering is present only in the SDW phase, once the SDW is removed, the magnetic order also disappears on a sub-100~fs time scale, as the sample transitions to the paramagnetic phase. We thus conclude that the SDW is directly driven by the transient $T_{e}$ and that even on ultra short time scales the transient heating of the electrons drives the spin-ordering of the SDW transition. The lattice plays a limited role on the SDW transition at these fluences, acting as a heat sink allowing the heated electron-spin gas to cool and the SDW ordering to re-emerge.

In summary, we have used trARPES at XUV energies to investigate the excitation and recovery of the SDW in Cr(110) thin films on ultrafast time scales. We find that the transition to the paramagnetic state occurs promptly within the pump-pulse duration. In addition, we show that the order parameter $\Delta$ follows the electronic temperature $T_{e}$ which governs both the closing and re-opening of the SDW gap. This therefore suggests that the electrons form a quasi-equilibrium with the spin order, while the phonon temperature lags behind. We speculate that such a general approach may be applicable to a range of optically induced phase transitions in the ultrafast regime, and may be used as the basis of more sophisticated descriptions including lattice or other degrees of freedom.

\begin{acknowledgments}
C.M. acknowledges support by the Swiss National Science Foundation under grant number PZ00P2\_154867, as well as by the Alexander von Humboldt Foundation

\end{acknowledgments}

\newpage

\section{Supplementary Material}

\subsection{Two Temperature Model}

In the main article the transient electronic temperature $T_{e}$ was extracted from the broadening of the Fermi level. By using a two-temperature model (2TM), it is possible to model this transient temperature and additionally estimate the temperature of the lattice. The equations for the 2TM are two coupled differential equations:

\[C_{e}\frac{dT_{e}}{dt} = P\sqrt{\frac{2.77}{\pi\sigma^{2}}}\exp \left(\frac{-~2.77~(t-t_{0})^{2}}{\sigma^{2}}\right) - \lambda (T_{e}-T_{latt})
\]

\[C_{latt} \frac{dT_{latt}}{dt} = \lambda(T_{e}-T_{latt})
\]

$C_{e}=\gamma T_{e}$ and $C_{latt}$ are the electron and lattice specific heat capacities; P is the absorbed energy density; $\sigma$ is the full-width (in time) at half maximum of the laser pulse; and $\lambda$ is the electron-phonon coupling. The factor of 2.77 comes from using the full width at half maximum of the Gaussian pulse.
The values for P~=~75~J~cm$^{-3}$ and $\sigma$~=~65~fs are determined experimentally; $\gamma$~=~75~J~m$^{-3}$~K$^{-2}$ is calculated within the Sommerfeld model \cite{Ashcroft1976}; and the adjustable fit parameter $\lambda$ is set to 0.22.
The temperature dependence of $C_{latt}$ is approximated in the Debye model, with a high temperature value of $C_{latt}$ = 0.448~J~g$^{-1}$~K$^{-1}$ \cite{Haynes2003}. The equations are solved numerically and reproduce the experimental data.

\subsection{Details of Mean-Field Model}

To model the transient response of the spin density wave (SDW) we have used a simple model describing one band of electrons with a linear dispersion subject to an interaction described in a mean-field manner by an order parameter. Such a theory is applicable to superconductors and spin and charge density waves, as it gives a mean field description of the relevant interaction, which need not be explicitly specified but is contained in the order parameter \cite{Gruner2000,Mahan2000}. Depending on the exact interactions between electrons and holes in a material the ground state which develops can be a superconductor (singlet or triplet state), SDW or CDW. The effect of the order parameter is then to open a gap of magnitude $\Delta$ in the linear dispersion.

The spectral function \textit{A(k)} in this model has the form \cite{Mahan2000}:

\[A(k,\omega)=u(k)^{2}\delta(\omega-E_{+}(k))+v(k)^{2}\delta(\omega+E_{-}(k))
\]

where $u(k)$ and $v(k)$ are the coherence factors and $E(k)$ is the renormalised dispersion in the SDW state. The renormalised dispersion is given by:

\[E_{\pm}(k) = \pm \sqrt{\epsilon(k)^{2}+\Delta^{2}}
\]

$\epsilon(k)$ is the bare dispersion and $\Delta$ is the order parameter, which is equal to half the gap size measured by photoemission. In the gapped state, two renormalised dispersion branches appear, which carry spectral weight corresponding to  $u(k)^{2}=(1+\epsilon/E_{\pm})/2$ and $v(k)^{2}=(1-\epsilon/E_{\pm})/2$
respectively. $u(k)^{2}$ and $v(k)^{2}$ are the so-called coherence factors, which respectively describe the spectral weight carried by the renormalized dispersions $E_{+}$ and $E_{-}$ in the SDW phase. As such, they describe the spectral weight transferred from the original bare band $\epsilon(k)$ to the back-folded band in the SDW, which we monitor in the time-domain in our present study.

\begin{figure}
		
		\centering
		\includegraphics[scale=0.45]{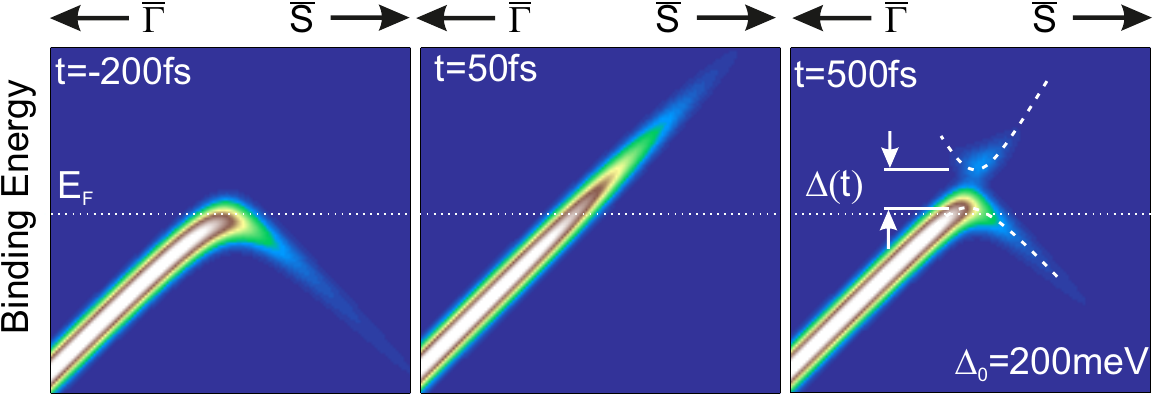}
		
	\caption{Simulated spectral function of the SDW in Cr before, shortly after and at longer time after excitation by a laser pulse. Changes are due to dynamics of the order parameter and the broadening of the Fermi distribution.} 
	\label{fig:Figure 5}

\end{figure}

We model the dynamics of the SDW phase by utilising the above spectral function. By defining $\Delta=\Delta(T_{e})$ we can simulate the evolution of the spectral function following laser excitation. Images from the simulation are shown in Fig.~5 for three representative delays: before excitation, shortly after excitation, and at later delays. A Fermi distribution function has been applied here to model the population of the upper and lower bands, in contrast to Fig.~4a in the main text. Before excitation, the back folded band due to the renormalised dispersion in the SDW phase is evident. Upon excitation this changes  into the paramagnetic high-temperature phase as $\Delta$ goes rapidly to zero, removing spectral weight from the back-folded dispersion. After some time $\Delta$ becomes finite again, and the gap gradually re-opens, leading to a recovery of the back-folded spectral weight. Intensity is found in both branches of the renormalised dispersion as electrons continue to relax.

\begin{figure}

		\centering
		\includegraphics[scale=0.3]{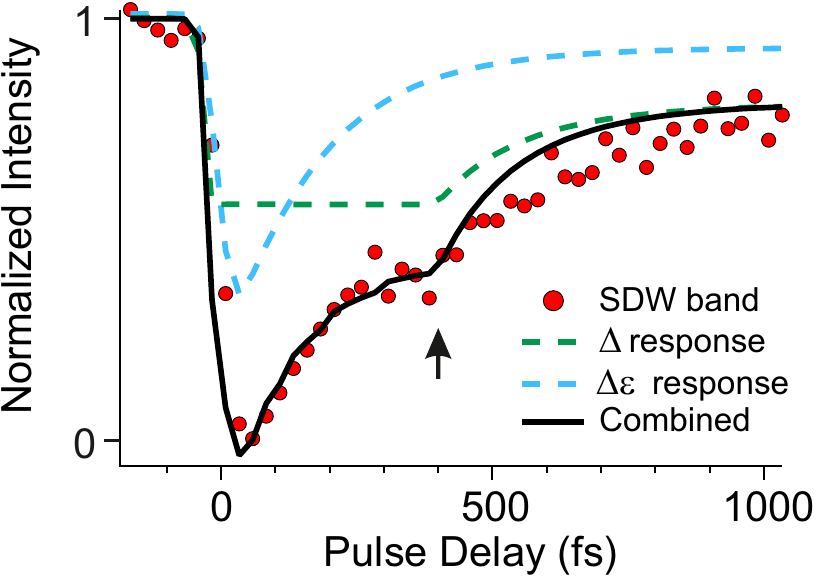}
		
	\caption{Dynamics of the normalized SDW band intensity due to the order parameter dynamics (green dashed line) and rigid shift (blue dashed line), and their combination (solid black line) compared with the experimental data (red markers).} 
	\label{fig:Figure 6}

\end{figure}

To further illustrate the model, Fig.~6 shows the change of intensity within a fixed region of interest caused by the SDW order parameter (green) and the thermal relaxation or rigid shift dynamics (blue) described in the main text. The transients are calculated independently for each contribution, highlighting the shape of the transient associated to each effect.

\subsection{Fluence-Dependent Response}

In the main article, it was shown that recovery of the SDW intensity is observed on two time scales. In Fig.~7 we show the same analysis for three additional absorbed fluences, in which similar dynamics in the recovery is observed for the two higher fluences. The lowest fluence shows only a single exponential decay. All curves shown here are for an excitation with p-polarised light.

\begin{figure}
		
		\centering
		\includegraphics[scale=0.3]{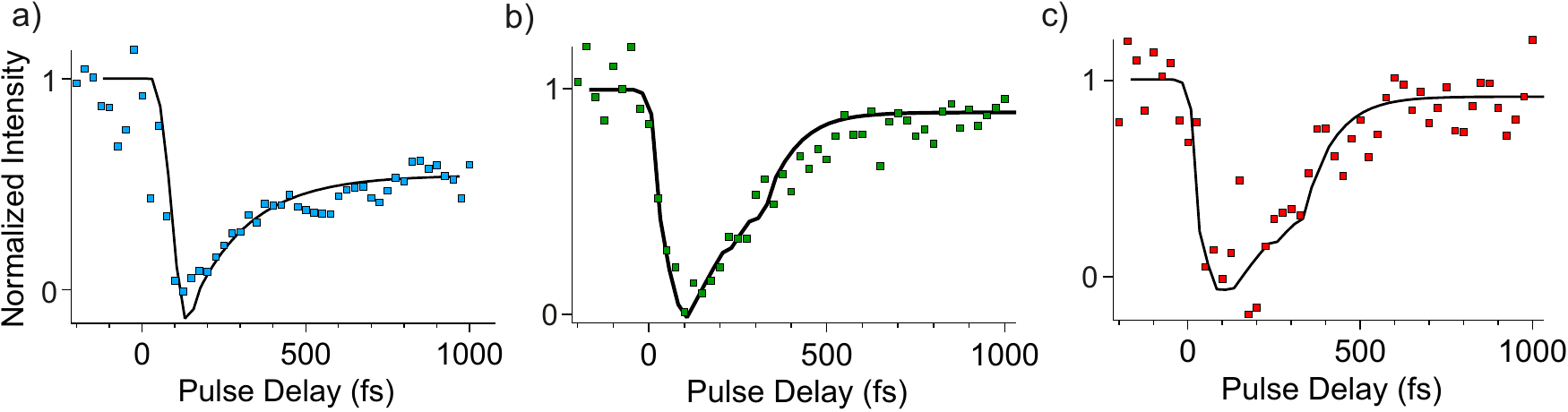}
		
	\caption{\small{Response of SDW band in renormalised region as in the main article for a) 0.17~mJ~cm$^{-2}$ b) 0.26~mJ~cm$^{-2}$ and c) 0.48~mJ~cm$^{-2}$. The corresponding simulated curves are shown in black. In b) and c) the points at which the SDW gaps re-open are 300~fs and 325~fs respectively, in comparison with 400~fs as shown in the main text for a fluence of 0.8~mJ~cm$^{-2}$.}} 
	\label{fig:Figure 7}

\end{figure}

\vspace{5mm}

\bibliography{../../../library}

\end{document}